\documentclass[%
 reprint,%
 amssymb, amsmath,%
 aip,cha,%
,twocolumn,superscriptaddress, floatfix]{revtex4-1}
\usepackage{amsmath,amssymb,graphicx,epstopdf,color,bm,soul,upgreek, mathtools}

\begin{document}

\title{Counter-directional polariton coupler}

\author{M. Klaas}%
\affiliation{Technische Physik, Wilhelm-Conrad-R\"ontgen-Research Center for Complex
Material Systems, Universit\"at W\"urzburg, Am Hubland, D-97074 W\"urzburg,
Germany}
\author{J. Beierlein}
\affiliation{Technische Physik, Wilhelm-Conrad-R\"ontgen-Research Center for Complex
Material Systems, Universit\"at W\"urzburg, Am Hubland, D-97074 W\"urzburg,
Germany}
\author{E. Rozas}
\affiliation{Dept. F\'{i}sica de Materiales \& Inst. N. Cabrera, Univ. Aut\'{o}noma, Madrid 28049, Spain}
\author{S. Klembt}%
\affiliation{Technische Physik, Wilhelm-Conrad-R\"ontgen-Research Center for Complex
Material Systems, Universit\"at W\"urzburg, Am Hubland, D-97074 W\"urzburg,
Germany}
\author{H. Suchomel}%
\affiliation{Technische Physik, Wilhelm-Conrad-R\"ontgen-Research Center for Complex
Material Systems, Universit\"at W\"urzburg, Am Hubland, D-97074 W\"urzburg,
Germany}
\author{T. H. Harder}%
\affiliation{Technische Physik, Wilhelm-Conrad-R\"ontgen-Research Center for Complex
Material Systems, Universit\"at W\"urzburg, Am Hubland, D-97074 W\"urzburg,
Germany}
\author{K. Winkler}%
\affiliation{Technische Physik, Wilhelm-Conrad-R\"ontgen-Research Center for Complex
Material Systems, Universit\"at W\"urzburg, Am Hubland, D-97074 W\"urzburg,
Germany}
\author{M. Emmerling}%
\affiliation{Technische Physik, Wilhelm-Conrad-R\"ontgen-Research Center for Complex
Material Systems, Universit\"at W\"urzburg, Am Hubland, D-97074 W\"urzburg,
Germany}
\author{H. Flayac} 
\affiliation{%
Institute of Physics, Ecole Polytechnique Federale de Lausanne (EPFL), CH-1015 Lausanne, Switzerland}
\author{M. D. Mart\'{i}n} 
\affiliation{Dept. F\'{i}sica de Materiales \& Inst. N. Cabrera, Univ. Aut\'{o}noma, Madrid 28049, Spain}
\author{L. Vi\~{n}a} 
\affiliation{Dept. F\'{i}sica de Materiales \& Inst. N. Cabrera, Univ. Aut\'{o}noma, Madrid 28049, Spain}
\affiliation{Instituto de F\'{i}sica de la Materia Condensada, Universidad Aut\'{o}noma, Madrid 28049, Spain}
\author{S. H\"ofling}
\affiliation{Technische Physik, Wilhelm-Conrad-R\"ontgen-Research Center for Complex
Material Systems, Universit\"at W\"urzburg, Am Hubland, D-97074 W\"urzburg,
Germany}
\affiliation{SUPA, School of Physics and Astronomy, University of St Andrews, St Andrews
KY16 9SS, United Kingdom}
\author{C. Schneider}
\affiliation{Technische Physik, Wilhelm-Conrad-R\"ontgen-Research Center for Complex
Material Systems, Universit\"at W\"urzburg, Am Hubland, D-97074 W\"urzburg,
Germany}

\begin{abstract}
{We report on an on-chip routing device for propagating condensates of exciton-polaritons. This counter-directional coupler implements signal control by a photonic microdisk potential, which couples two lithographically defined waveguides and reverses the condensate's propagation direction. By varying the structural sizes, we utilize the conjunction of the different dimensionalities to additionally evidence the functionality of a polaritonic resonant tunnel diode. Furthermore, we investigate the ultra fast dynamics of the device via ps-resolved streak camera measurements, which is distinctive for the polariton platform. This scalable, all-directional coupler element is a central building block for compact non-linear on-chip photonic architectures.}
\end{abstract}

\maketitle

\textit{Introduction.---} 

The operation frequency and power consumption of classical on-chip circuits are limited by the response of charged carriers to externally applied electric fields. Photonic on-chip technologies for information processing, in turn, promise ultra fast operation speed and low power consumption thanks to low losses. Hence, they carry the potential to overcome such intrinsic barriers. Yet, photons interact weakly with themselves and their environment \cite{Miller2010}, which makes the implementation of logic devices challenging. 

However, when photons strongly couple to matter excitations in a semiconductor microcavity \cite{Weisbuch.1992}, a strong non-linearity appears \cite{Vladimirova.2010}, which can be utilized to implement high-speed photonic building blocks for information processing. Furthermore, the continuous development of high quality microcavities \cite{Schneider.2016} has enabled the expansion of polariton condensates over macroscopic distances \cite{Nelsen2013}. 

To date, one has witnessed a variety of demonstrator experiments for on-chip steering and manipulation of the expansion of coherent polariton condensates, including condensates propagating along waveguides \cite{Wertz2012, Fischer.2014, Anton.2014}, switches, spin-filters, transistors and amplifiers and even polaritons populating a topologically protected edge mode \cite{Ballarini.2012, Gao.2012, Amo.2010, Sanvitto.2016, Niemitz.2016, Klembt.2018}. Based on the ultra fast switching dynamics of expanding polaritons signals, complete on-chip logic architectures have been proposed \cite{Liew.2008}.

One of the main future challenges in polariton logic is the realization of an all-electrical excitation, manipulation and readout scheme \cite{Tsintzos.2008}. To this end, electrical polariton condensation \cite{Schneider.2013, Bhattacharya.2013} and switching \cite{Dreismann.2016, Suchomel.2017} has been achieved. Combining electrical control with the recent research on room temperature condensation in GaN \cite{Christopoulos.2007}, organic materials \cite{Plumhof.2014, Daskalakis.2014, Dietrich.2016} and the prospects of transition metal dichalcogenides \cite{Schneider.2018} would offer a powerful, versatile and competitive platform. 

Photonic, and likewise, polaritonic optical networks require passive routing elements to fully harness the system as a low power consumption \cite{Deng.2003} and coherent on-chip architecture. Basic routing effects have been achieved and predicted for polaritons \cite{Flayac.2013, Marsault.2015, Nguyen.2013}, which show some functionalities but are mainly based on active optical control via a tunable auxiliary exciton reservoir. 

Here, we demonstrate a compact polariton counter-directional router which operates as a polaritonic resonant tunnel diode. The device can feasibly be scaled to large logic architectures without the requirement for active external control parameters. It relies on the engineering of the photonic potential so as to reverse the propagation direction of an input polariton flow.

\vspace{\baselineskip}
\textit{Technology and Characterization.---}

\begin{figure}
\includegraphics[width=\linewidth]{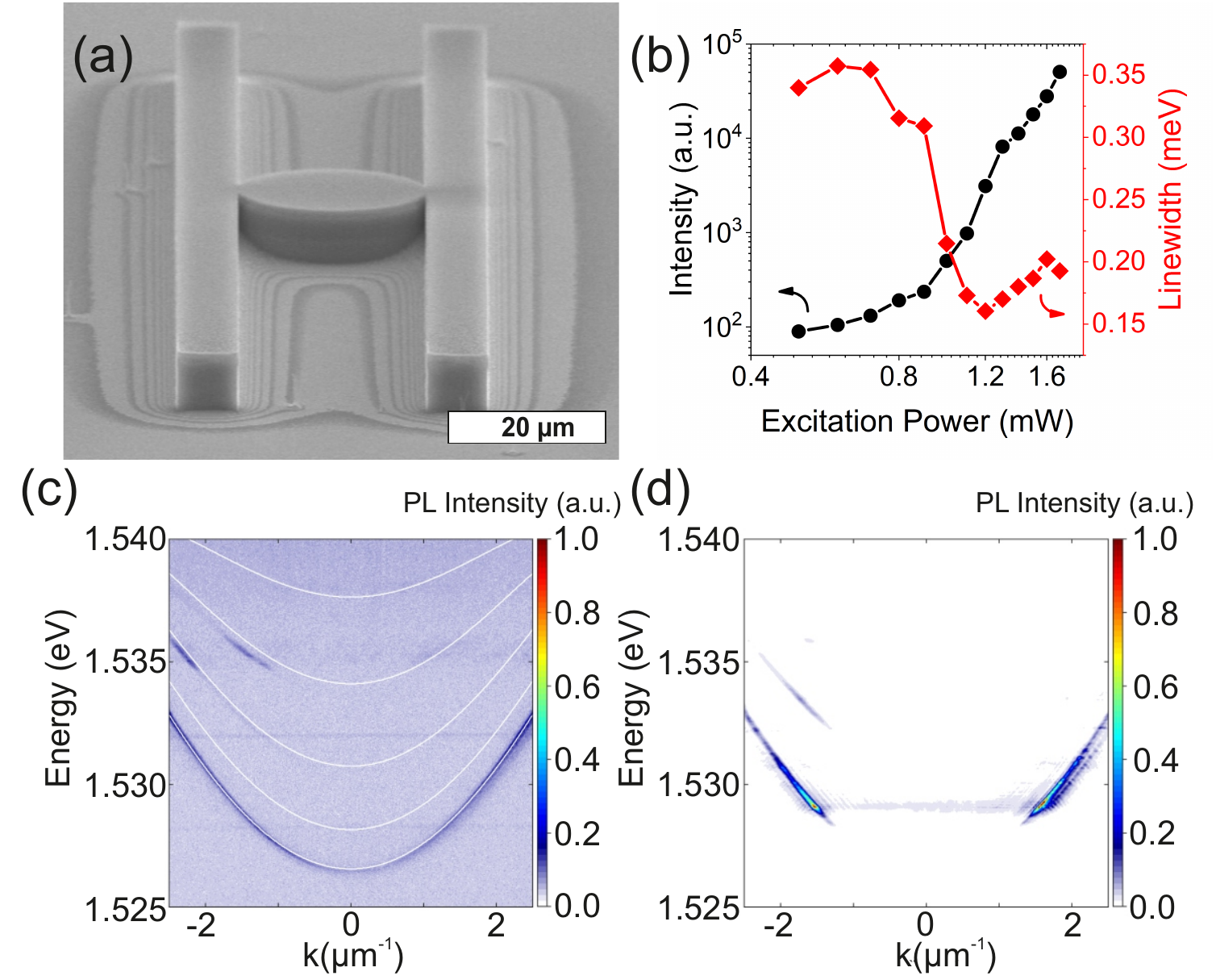} 
\caption{(a) SEM image of the microdisk device coupling two waveguides. (b) Input-output analysis of such a device extracted from the parameters of a Lorentz fit of the emission at k $\approx 1.7 \upmu m^{-1}$. An intensity nonlinearity (left axis, circles) and a coherence buildup, evidenced by a linewidth drop (right axis, diamonds), are oberved at the excitation power thershold (0.9 mW). (c) Dispersion relation in the linear regime with the subbands originating in the one-dimensional photonic confinement of the waveguides at an excitation power of 0.5 mW. (d) Momentum resolved condensate emission showing propagation along both directions of the waveguide, indicated by the selective population of distinct non-zero wavevectors at an input power of 1.5 mW.}
\label{fig1}
\end{figure}

The device has been processed on a sample grown by molecular beam epitaxy. The cavity consists of 36 AlAs/AlGaAs mirror pairs (bottom) and 32 mirror pairs (top) with three stacks of four 13 nm GaAs QWs in an AlGaAs Lambda-width cavity design. The Rabi splitting, which was determined by white light reflection measurements, is 9.4 meV. The Q factor is 12000, determined by a low power measurement of a photonic structure. The microdisk device was deeply etched (through the QW layer) to provide photonic confinement. Figure \ref{fig1} a) depicts a slightly tilted topview scanning electron microscopy (SEM) image of the device fabricated in such a manner. We produced several devices with waveguide widths ranging from nominally 2 to 4 $\upmu$m and microdisk diameters of 10-40 $\upmu$m.

The experimental measurements have been carried out with a photoluminescence setup capable of Fourier- and real space detection. Excitation was facilitated by a tuneable, mode locked Titanium Saphire laser providing 2 ps pulses, which were tuned to the first reflectivity minimum of the microcavity for efficient injection. The excitation is mechanically chopped with a ratio of 1:12 (f=8000rpm) to reduce sample heating. Furthermore, a tomography technique was implemented via motorized lenses to allow for energy selective real space imaging. The pump spot was focused using a microscope objective with NA = 0.42 to a diameter of $\approx 3$ $\upmu$m full width at half maximum. 

We show an example characterization of a waveguide coupled to a microdisk in Fig. \ref{fig1}. The measurements have been performed at 5 K on a straight part of a waveguide. Figure \ref{fig1} (b) shows the emission intensity (left axis, circles) obtained by a Lorentzian fit at k $\approx 1.7 \mu m^{-1}$ as a function of the excitation power. It exhibits a characteristic intensity nonlinearity and linewidth drop (right axis, diamonds) at the threshold, which can be related to coherence buildup and therefore polariton condensation. The population of distinct non-zero wavevectors reflects the propagation along the waveguide structures in both directions. Figure \ref{fig1} (c) presents a low power measurement (0.5 mW) of the dispersion relation evidencing photoluminescence from several subbands, characteristic for the waveguides' one dimensional confinement. These subbands have been fitted by a square well photonic potential \cite{Tartakovskii.1998}, yielding a waveguide width of 4.5 $\upmu$m [in good agreement with the width shown in Fig. \ref{fig1} (a)]. In addition, it allows to extract the detuning of -16 meV with an exciton energy of 1.543 eV.  At a threshold excitation power ($\approx 0.9$ mW), a polariton condensate is formed which propagates along both directions of the waveguide with a momentum of k $\approx 1.7 \mu m^{-1}$ [see Fig. \ref{fig1} (d)].

\vspace{\baselineskip}
\textit{Experimental Results, Modelling and discussion.---}

\begin{figure}
\includegraphics[width=\linewidth]{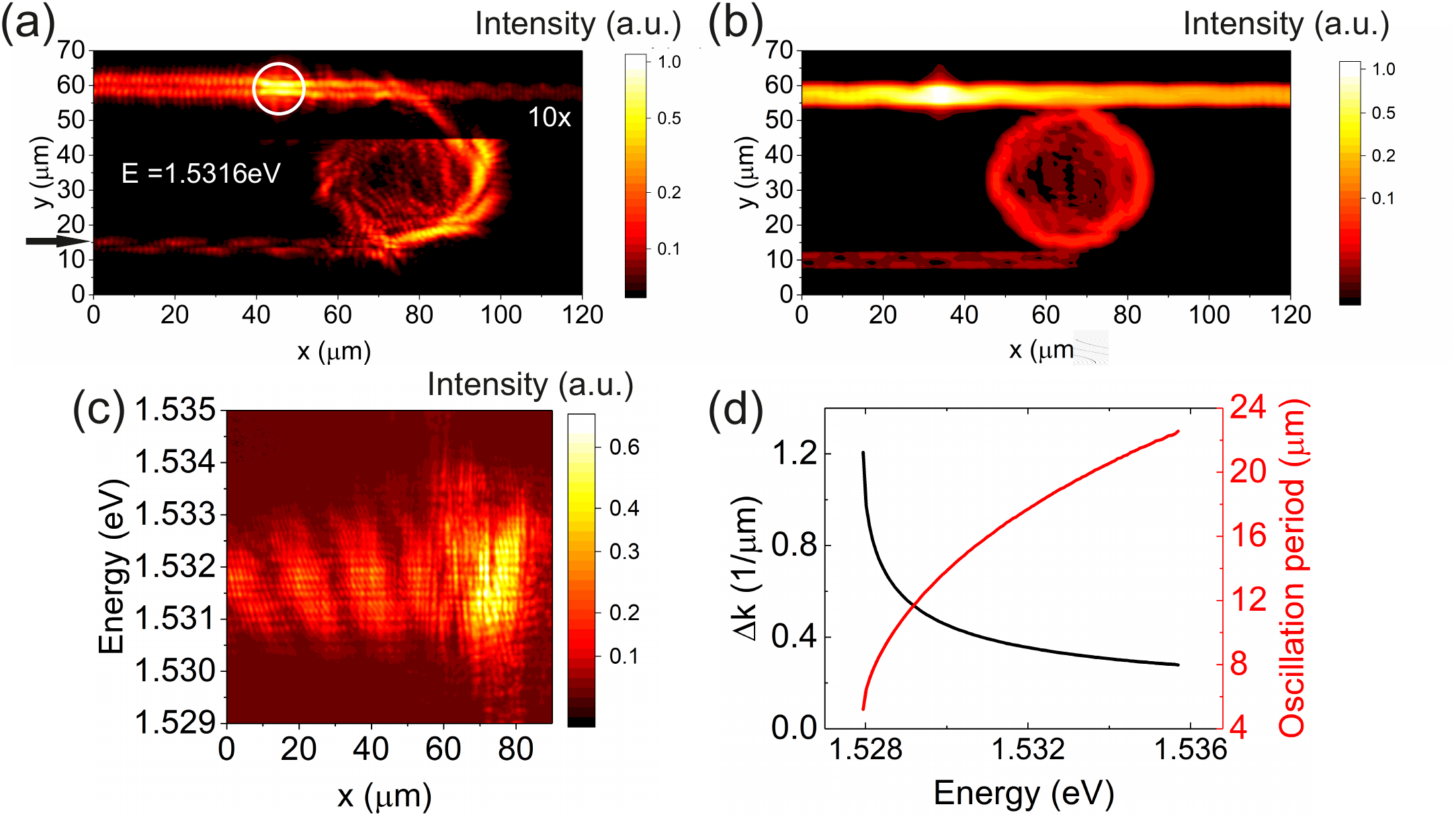} 
\caption{(a) Intensity color-coded real space image of the condensate expanding through the microdisk device. A neutral density filter (factor of 10) was present for the region y$>$45. (b) Simulated plot of the intensity distribution and mode pattern upon excitation at a similiar position. (c) Energy- and real space-resolved signal extracted at the position of the arrow in (a) at approx. y=15 $\upmu$m. A clear tilt of the fringes is observed. (d) Theoretical oscillation frequency and period dependent on energy.}
\label{fig2}
\end{figure}

We now turn to an investigation of the functionalities of a microdisk device with structure parameters of 4 $\upmu$m and 40 $\upmu$m for the waveguide and disk width, respectively. This configuration serves as a counter-directional optical coupler. Fig. \ref{fig2} (a) shows a real space image of the polariton flow (excitation power in the condensate regime at 1.3 mW). This flow is steered by the photonic potential created by the disk structure with the excitation spot at the position of the white circle. The polaritons are guided into the microdisk by the waveguide and rotate inside due to internal reflection such that their propagation reversed in the output guide. This confirms the counter-coupler capability of the structure. The effect can be observed from every entrance direction, routing exactly to the opposite waveguide with reversed propagation momentum. The intensity of the top part of the image is reduced by a factor of 10 by means of a neutral density filter and the real space image is resolved in energy (E = 1.5316 $\pm$ 0.0001  eV) to improve the visibility of the propagating mode. We calculate the coupling efficiency of the device to $\approx$ 11.5\% from the ratio between the emission just after the polaritons' entrance into the microdisk and that just before their exit from the structure.

The propagation can be modeled analogously to Ref. \cite{Winkler.2017} at a mean field level by means of a 2D driven-dissipative Ginzburg-Landau equation for the polariton wave function:
\begin{equation}
\begin{aligned}
(i- \eta)\hbar \frac{\partial \Psi}{\partial t} =  &- \frac{\hbar^2 \Delta}{2m}\Psi + U(x) \Psi + \alpha \lvert \Psi \rvert^2 \Psi \\
&+\frac{i \hbar}{ 2}\left(P-\gamma - \Gamma \lvert \Psi \rvert^2 \right)\Psi.
\end{aligned}
\end{equation}
Using the following parameters: $\alpha=10^{-3}$meV$\upmu$m, loss rate $\gamma=5*10^{-2}$ps$^{-1}$, relaxation rate $\eta=10^{-2}\gamma$ and with the excitation profile $P(x)=50 \gamma e^{\frac{-(x-x_0)^2}{dx^2}} e^{\frac{-(y-y_0)^2}{dy^2}}$ (dx=dy=2.5$\upmu$m) with $\Gamma=0.1 \gamma$. $U(x)$ is the potential landscape, originating in the photonic confinement of the etched structures, the potential was taken as the nominal layout of the etched structure, waveguide width=4 $\upmu$m and disk width=40 $\upmu$m. The result of this model is depicted in Fig. \ref{fig2} (b) for a simulation of the whole device with good qualitative agreement with the experiment. The slight discrepancies stem from the dominance of the first excited state in the experiment (and the faster decay rate, however, we note that the simulated intensity distribution at the microdisk and exit waveguide, which are the focus of this study, match well with the experimental one.

Interestingly, at the exit waveguide, we observe a mode beating pattern which is created by the disk resonantly feeding a superposition of two different waveguide modes in the outcoupling region. In order to investigate the modes at the exit waveguide in a more detailed manner with respect to their propagation, Figure \ref{fig2} (c) depicts the energy-resolved emission as a function of the position in the outcoupling region \ref{fig2} (a) (a cut through the mode indicated by the black arrow in \ref{fig2} (a). We observe a tilted pattern of the emission which corresponds to a different period/frequency of the oscillation at different propagation energies. This is a consequence of the changing wavevectors of the competing modes along their dispersion. Fig. \ref{fig2} (d) plots the theoretical oscillation period as a function of propagation energy, extracted from a fit of the two polariton branches of the first and second mode. It also shows a tilt of the oscillation period with different propagation energy. We note, that the observed effect is only visible for a waveguide wide enough such that the energy of the second transverse mode inside the waveguide is at an energy fed by the microdisk.

\begin{figure}
\includegraphics[width=\linewidth]{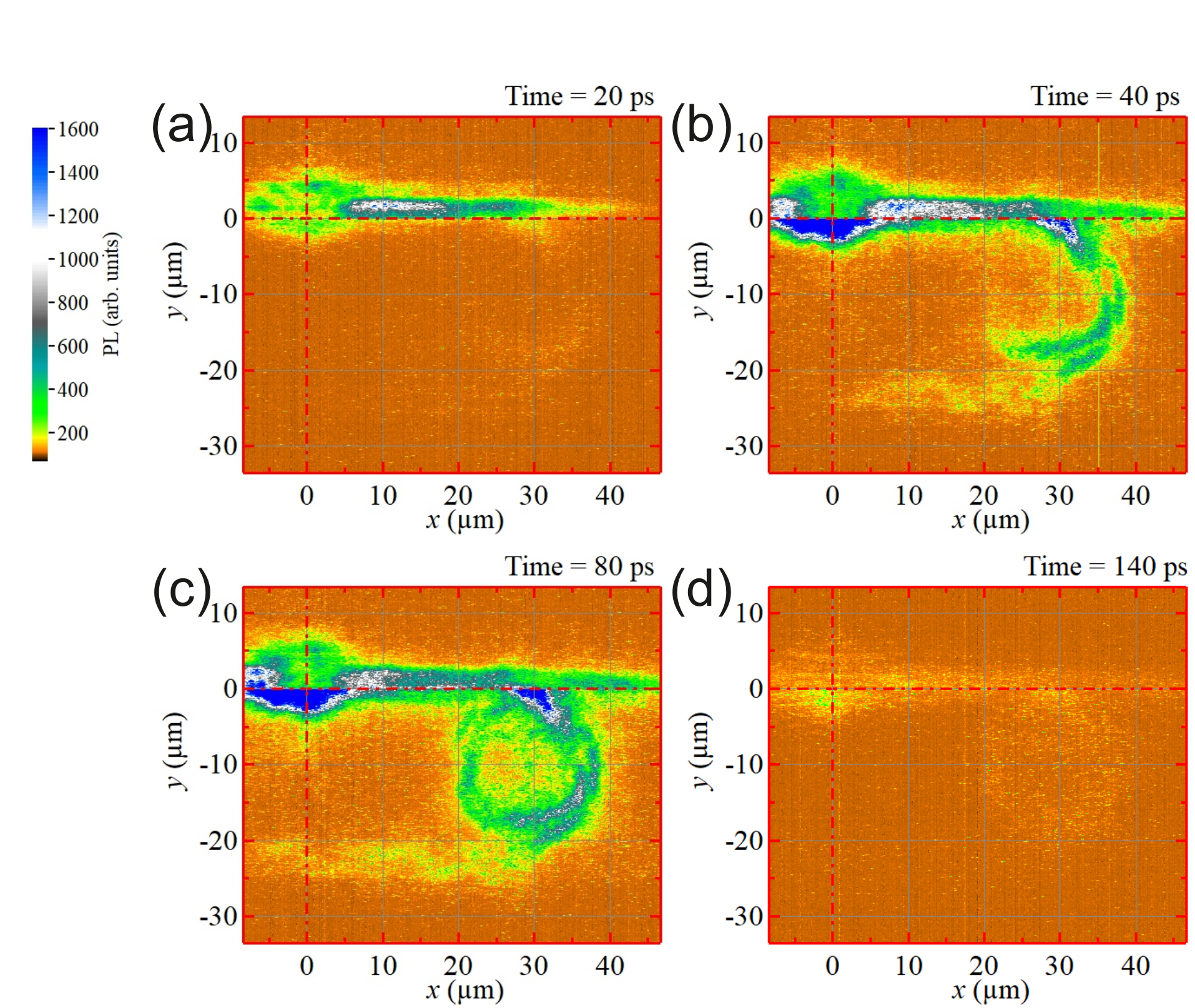} 
\caption{Time-resolved series of the polariton propagation along a microdisk device, the horizontal dashed red line marks a change in gain of the streak camera (factor of 3). (a) Spatial emission distribution at t=20 ps, the polaritons have started propagating along the entrance waveguide. (b) at 40 ps the emission reveals that polaritons have arrived at and crossed through the microdisk to the other adjacent waveguide. (c) At 80 ps, the oscillations have manifested at the exit port and the disk shows a circulating mode pattern. (d) At 140 ps most of the polaritons have decayed and the dynamics has ended.}
\label{fig3}
\end{figure}

A core property of exciton-polaritons is their ultrafast dynamics while still possessing matter characteristics. To evidence and analyze this, we employ a streak camera measurement (resolution approx. 10 ps) to trace the condensate flow of a microdisk device in time. Fig. \ref{fig3} shows a series of such measurements at different time delays to highlight the important points of the dynamics. In \ref{fig3} (a), approximately 20 ps after the arrival of the excitation pulse, the condensed polaritons have started to move along the waveguide. At t=40 ps (panel \ref{fig3} (b)) the flow begins to cross into the second waveguide via the microdisk structure. At 80 ps in \ref{fig3} (c) the oscillations have completely manifested and at 140 ps the polariton dynamics inside the device have ceased due to their finite lifetime on the order of 39 $\pm$ 5 ps outside of the reservoir. This is comparable to the findings with etched waveguide structures in the literature where lifetimes of 17.5 ps \cite{Marsault.2015} and 18 ps \cite{Gao.2012, Anton.2012} have been reported, with propagation lengths on the order of 100 $\upmu$m. 

\begin{figure}
\includegraphics[width=\linewidth]{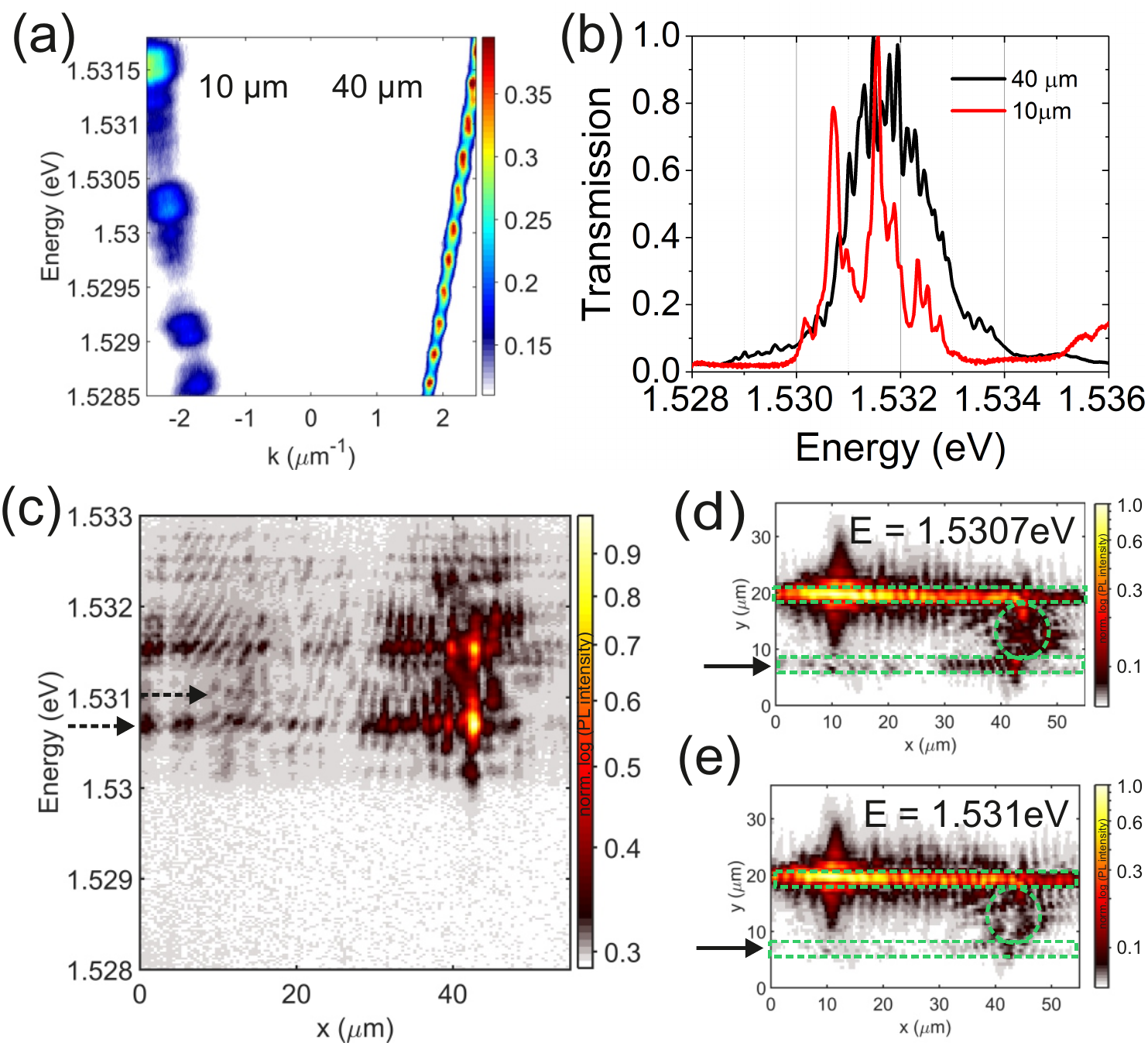} 
\caption{(a) Angle-resolved, discretized mode structure of microdisks with a diameter of 10 $\upmu$m and 40 $\upmu$m under low power excitation, where, for the sake of comparison, only negative (positive) $k_y$ values are depicted for the 10 $\upmu$m (40 $\upmu$m) disk. (b) Energy-resolved emission of the condensate exiting the microdisks filtered in real space directly at the junction of the microdisk and the waveguide. (c) Emission along the exit waveguide resolved in energy for the 10 $\upmu$m structure. (d) Energy resolved real space emission (1.5307 eV) of the exit-waveguide, depicting propagation of one mode through the disk into the bottom waveguide. The device layout is indicated by the green dashed lines. (e) Shows filtered out energy range of a previously propagating mode (1.531 eV).}
\label{fig4}
\end{figure}

Now we study the effects on the device operation of certain parameters of the coupler, such as the waveguide width and microdisk diameter. For smaller diameters of the disk and waveguide section, it is possible to find combinations which additionally use the routing device as a filter for specific modes (thus producing a monochromatic signal propagating along the exit port). This occurs due to the necessity for energy matching at the junction between the quasi zero-dimensional confinement of the disk and the one-dimensional confinement of the waveguide. This layout therefore constitutes a resonant-tunnel barrier for polariton condensates with the additional counter-directional routing already evidenced. The disk diameter is chosen such that a high degree of discretization is observed in the mode structure to allow fine selection of the propagation energy. Fig. \ref{fig4} (a) shows a typical comparison of the mode structure at low excitation power of two different investigated microdisks (10/40 $\upmu$m); having in mind that under these conditions the emission is symmetric in $k$, for the sake of comparison, only negative (positive) $k_y$ values are depicted for the two disks. The desired stronger discretization of the photonic modes inside the traps is clearly visible for the smaller diameter case. 

Figure \ref{fig4} (b) shows a polariton condensate (with a pumping power moderately above threshold) for both diameter structures energetically resolved and filtered directly in real space at the exit port of the microdisk. The nearly continuous emssion spectrum of the 40 $\upmu$m microdisk, with only slight discretization visible, as compared to the discrete spectum of the 10 $\upmu$m one, clearly evidences the changes in transmission due to confinement effects in the microdisks. The 10 $\upmu$m case exhibits two prominent levels at 1.5307 and 1.5315 eV, while the whole emission spectrum is shifted to lower energies due to slightly different detunings between both devices. This translates into the real space emission behaviour summarized in Fig. 4 (c), displaying the propagation energy as a function of spatial position along the exit waveguide, for the case of a 10 $\upmu$m microdisk coupled to a 2 $\upmu$m waveguide. Propagation can only be observed in specific modes, while others are blocked due to the mode structure imposed by the photonic confinements. Real space images in panel (d) and (e) each select specific energies 1.531 eV and 1.5307 eV via tomography and show the complementary configurations, demonstrating the control of the polariton propagation by selecting the appropriate energy, so there can be either propagation [\ref{fig4} (d)] or blockade [\ref{fig4} (e)]. This structure therefore enables mode selection in combination with flow direction manipulation for a polariton condensate, all without any external control parameter.

\vspace{\baselineskip}
\textit{Conclusions.---}

We have demonstrated the possibility for a passive and ultra fast polariton counter-directional router based on an arrangement of zero- and one-dimensional photonic confinements. Additionally, by engineering the size parameters and utilizing the specific propagating modes, the device also works as a filter that allows an energy selection obtaining a monochromatic exit signal. Furthermore, the configuration is easily scalable and integratable into polariton based logic networks. We evidenced this by a series of photoluminescence measurements involving a tomography and streak camera method, laying out the devices capabilities. Such detailed tailoring of the flow of these quantum fluids of light paves the way to harness their non-linearity in next generation photonics, for example in the implementation of topologically non-trivial polariton networks based on coupled disks and waveguides \cite{Kozin.2018}.

\begin{acknowledgments}
The W\"urzburg group acknowledges the financial support by the state of Bavaria and the DFG within the project Schn1376-3.1. J.B. and S.K. acknowledge funding from DFG grant KL3124/2-1. The Madrid team acknowledges financial support by the Spanish MINECO Grants MAT2014-53119-C2-1-R and MAT2017-83722-R. E.R. acknowledges financial support from a Spanish FPI Scholarship No. BES-2015-074708.
\end{acknowledgments}

\end{document}